# Do Galactic Systems Form Too Late in Cold+Hot Dark Matter Models?

Chung-Pei Ma

Theoretical Astrophysics 130-33, California Institute of Technology, Pasadena, CA 91125

and

Edmund Bertschinger

Department of Physics 6-207, Massachusetts Institute of Technology, Cambridge, MA 02139

## ABSTRACT

The abundance of galactic systems at high redshifts can impose a strong constraint on the cold+hot dark matter (CDM+HDM) models. The hot component reduces the excessive small-scale power in the COBE-normalized CDM model but also delays the epoch of galaxy formation. We present results from the first numerical simulations that have enough dynamic range to address accurately the issue of high-redshift halo abundances in CDM+HDM models. Equivalent high-resolution particle-particle/particle-mesh $N$-body simulations are performed for spatially flat models with $\Omega_\nu = 0.3$ and 0.2 (with $H_0 = 50$ km s$^{-1}$ Mpc$^{-1}$ and $\Omega_b = 0.05$). We study the constraints placed on the models by the high-redshift quasar space density and by the mass fraction in neutral dense gas associated with damped Ly$\alpha$ systems. We find that even with optimistic assumptions, the much-studied $\Omega_\nu = 0.3$ model does not produce enough massive halos to account for the observed abundance of quasars at $z > 4$. The model passes this test if $\Omega_\nu$ is decreased to 0.2. Both models do not produce enough high column-density halos to account for the amount of gas in damped Ly$\alpha$ systems at $z \gtrsim 3$: the $\Omega_\nu = 0.3$ model falls short by a factor $\sim 80$; the $\Omega_\nu = 0.2$ model by a factor $\sim 3$. We conclude that only CDM+HDM models with $\Omega_\nu \lesssim 0.2$ can match observations at high redshift, implying an upper bound of 4.7 eV on the most massive light neutrino (presumably the $\tau$).

astro-ph/9407085  26 Jul 94

## 1. Introduction

The cold+hot dark matter (CDM+HDM) model of structure formation has emerged as a promising alternative to the standard CDM model. Replacing some of the CDM with HDM (massive neutrinos) suppresses small-scale power for a fixed large-scale normalization such as that provided by the cosmic microwave background anisotropy. This suppression helps to alleviate the problem of excessive small-scale power in the CDM model (e.g., Gelb & Bertschinger 1994) but it also delays the epoch of galaxy formation. Thus, the abundance of high-redshift quasars and the amount of mass in dense gas clouds can be used to constrain the amount of HDM present.

Early works based on linear calculations (Schaefer, Shafi, & Stecker 1989; van Dalen & Schaefer 1992; Taylor & Rowan-Robinson 1992; Holtzman & Primack 1993) and $N$-body simulations (Davis, Summers, & Schlegel 1992; Klypin et al. 1993; Jing et al. 1994) have shown general agreement between observations at redshifts $z \approx 0$ and the $\Omega_\nu = 0.3$ flat CDM+HDM model. Some of these workers have expressed concern about late galaxy formation, but the lack of adequate dynamic range in mass and length in the fully nonlinear regime prevented them from reaching firm conclusions.

Analytical approximations have been used by many workers to avoid the limited dynamic range of numerical simulations. The abundance of bound objects at high redshifts has been estimated using the Press-Schechter (Kashlinsky 1993; Haehnelt 1993) and the Gaussian peak (Nusser & Silk 1993) formalism in the hope of constraining various dark-matter models using quasar statistics. Similar calculations have been made comparing analytic models with the damped Ly$\alpha$ absorption systems (Mo & Miralda-Escude 1994; Kauffmann & Charlot 1994; Subramanian & Padmanabhan 1994). However, the accuracy of these approximations in the appropriate regime is uncertain. The peak model for halo formation, for example, has been criticized recently by Katz, Quinn, & Gelb (1993). As we will show, the Press-Schechter formula also can lead to inaccurate results for rare objects if not properly calibrated.

To be sure, uncertainties remain in various aspects of both the theories and observations of quasars and Ly$\alpha$ systems. Our understanding of the relation between the observed quasar activity on parsec scales and the host galaxies of size $\sim 10$ kpc is incomplete. Quasar formation itself cannot be addressed directly by any current cosmological simulations, but progress has been made in modeling the formation and fueling of black holes in the host galaxies of quasars. Efstathiou & Rees (1988) observed that the epoch of galaxy formation in the biased ($b \approx 2.5$) CDM model is consistent with the observed luminous quasar density between $z = 2$ and 4, but a sharp cut-off at $z > 5$ is predicted. Turner (1991) argued that there is a considerable time lag between the formation of virialized halos and the birth of high-redshift quasars, pushing the halo formation epoch to $z > 5$ or possibly $z > 10$. A contrary view was recently expressed by Haehnelt & Rees (1993), who argued for a short delay between halo collapse and formation of a massive black hole.

Uncertainties also exist on the observational front in estimating the comoving number density of quasars at $z > 2$. The Palomar CCD grism survey (Schmidt, Schneider, & Gunn 1991, 1994) has yielded a decline in the space density of quasars with $M_B < -26$ by a factor $\sim 3$ per unit redshift for $z \gtrsim 2.5$. Multicolor photographic surveys, on the other hand, have shown a slightly steeper decline in one case (Warren, Hewett, & Osmer 1994) and are consistent with a constant comoving density for $M_B \lesssim -28$ at $2 < z < 4$ in another (Irwin, McMahon, & Hazard 1991).

Damped Ly$\alpha$ absorption systems offer an alternative measure of early galaxy formation. Whether these systems are associated with gas-rich dwarf galaxies or normal disk spirals is not absolutely clear. However, recent observations of QSO H0836+113 (Wolfe et al. 1992) have failed to confirm the detection of a narrow emission feature (Hunstead et al. 1990) which had been regarded as evidence for a star-forming dwarf. Wolfe et al. (1992) instead detected extended line emission from a large area of $96 \times 44\ h_{50}^{-2}$ kpc$^2$ in the same system. Current surveys appear to favor the association of high column density Ly$\alpha$ absorbers with normal disk galaxies instead of dwarfs (e.g., Steidel & Hamilton 1992, 1993).

Assuming that the damped Ly$\alpha$ systems are mainly dwarf galaxies, Klypin et al. (1994) concluded that the CDM+HDM models are viable, in contrast with the analytic results of Mo & Miralda-Escude (1994), Kauffmann & Charlot (1994), and Subramanian & Padmanabhan (1994). Instead of relying on analytic approximations, we will estimate directly from large cosmological $N$-body simulations the fraction of the mass with hydrogen column density above the survey completeness threshold and compare with observations. We will also compare the abundance of dense dark matter halos capable of hosting quasars with the observed quasar abundance.

## 2. Simulations and Halo Properties

We have performed high-resolution CDM+HDM gravitational $N$-body simulations using the adaptive particle-particle/particle-mesh code (P$^3$M)$^2$ (Bertschinger & Gelb 1991; Gelb & Bertschinger 1994). In all cases $\Omega_\nu + \Omega_c + \Omega_b = 1$ and $H_0 = 50$ km s$^{-1}$ Mpc$^{-1}$. Two different neutrino masses were considered: $m_\nu = 7.0$ eV ($\Omega_\nu = 0.3$) and $m_\nu = 4.7$ eV ($\Omega_\nu = 0.2$). The baryon fraction used in computing the initial conditions was set to the value implied by primordial nucleosynthesis, $\Omega_b = 0.05$ (Walker et al. 1991). In the $N$-body simulations the baryons were treated identically as CDM.

The initial conditions for the hot and cold particles were generated using the Monte Carlo procedure described in Ma & Bertschinger (1994a). Our method differs from that used by other workers (Davis et al. 1992; Klypin et al. 1993; Cen & Ostriker 1993; Jing et al. 1994) in that the initial positions and momenta of our HDM particles were sampled from the full perturbed neutrino phase space distribution, while their HDM momenta were drawn from the Fermi-Dirac distribution which is the zeroth-order approximation to the full distribution function. We also took into account the wavenumber dependence of the suppression of the CDM velocities owing to neutrino damping in these models (Ma & Bertschinger 1994a). The nonlinear calculations were begun at a redshift $z = 13.6$ ($\Omega_\nu = 0.3$) and $z = 21.5$ ($\Omega_\nu = 0.2$), when the rms density fluctuation in the CDM component was about 0.2.

The simulations of the two models were performed with identical parameters (aside from the neutrino masses) and the same realization of the Gaussian random field. The primeval fluctuations are assumed scale-invariant and isentropic, normalized to a COBE rms quadrupole moment $Q_{\rm rms-PS} = 17.6$ $\mu$K (Bennett et al. 1994). The simulation volume is a cube of comoving length 100 Mpc and the interparticle force law is a Plummer law with a comoving softening length of 50 kpc (or a proper softening length of 10 $h_{50}^{-1}$ kpc at $z = 4$). We used $128^3$ simulation particles to represent the CDM and 10 times as many HDM particles for an accurate sampling of the neutrino phase space. For the $\Omega_\nu = 0.2$ case the particle masses are $2.5 \times 10^{10}$ $M_\odot$ for CDM and $6.6 \times 10^8$ $M_\odot$ for HDM. The simulations evolved to $z = 1.5$ required a total of 1300 CPU hours on a Convex C-3880.

The dark matter halos in our simulations are identified with an improved version of the Denmax algorithm described in Gelb & Bertschinger (1994). At the redshift of interest, Denmax finds all the particles within closed density contours around each density peak by moving the particles along the gradient of the mass density field (defined using a Gaussian of comoving radius 100 kpc) until they settle into the peaks. This method identifies distinct density concentrations more reliably than the standard friends-of-friends algorithm. After identifying the particles associated with each peak we compute the energy of the particles in each halo and remove the unbound ones. CDM and HDM particles are treated identically.

Because the task of Denmax is to associate particles with density peaks regardless of the value of the density maximum, some of the halos identified by Denmax at high redshifts have not yet collapsed and virialized. We therefore impose an overdensity constraint of $\delta\rho/\rho \geq 200$ to select virialized halos. The effect of this cut is to reduce the mass of all halos and to eliminate small density enhancements that have not collapsed. The halo abundance is sensitive to the overdensity constraint. We find up to 5 times more halos when we apply the less stringent constraint $\delta\rho/\rho \geq 50$ used by Klypin et al. (1993; see Fig. 2a below). The higher overdensity constraint is appropriate for virialized halos.

The comoving number density of Denmax halos more massive than $M$, $N(M, z)$, is shown in Figure 1 for the two models. It is a very sensitive function of $\Omega_\nu$ and $\Omega_c$ at high redshifts. The results at $z > 4$ are not plotted for $\Omega_\nu = 0.3$ because the slower growth in this model produces little structure at this early time, and it is difficult to identify halos reliably. In contrast, virialized halos with $M \sim 10^{11.5} M_\odot$ are present as early as $z \approx 7.4$ in our $100^3$ Mpc$^3$ simulation box for $\Omega_\nu = 0.2$. In fact, the halo abundance at $z = 7.4$ in the $\Omega_\nu = 0.2$ model is nearly identical to that at $z = 4.0$ in the $\Omega_\nu = 0.3$ model.

We can calibrate and check the validity of the Press-Schechter approximation (Press &

Schechter 1974) for the number density of virialized mass lumps:

$$N(M,z) = \int_{\ln M}^{\infty} d\ln M' \sqrt{\frac{2}{\pi}} \frac{\rho_0(z)}{M'} \frac{\delta_c}{\sigma(M',z)} \left|\frac{d\ln\sigma}{d\ln M'}\right| \exp\left[-\frac{\delta_c^2}{2\sigma^2(M',z)}\right], \qquad (1)$$

where $\rho_0(z)$ denotes the background mass density of the universe, $\delta_c$ is a free parameter characterizing the linear overdensity at the onset of gravitational collapse, and $\sigma^2(M,z)$ is the variance of the density field at redshift $z$, computed using the linear power spectrum $P(k,z)$ and a smoothing window function $W(kR)$. The power spectrum used here is taken to be the density-averaged $P = [\Omega_\nu P_\nu^{1/2} + (\Omega_c + \Omega_b)P_c^{1/2}]^2$, where $P_\nu$ and $P_c$ are computed numerically from the coupled, linearized Boltzmann, fluid, and Einstein equations that govern the evolution of the CDM and HDM density fields (e.g., Ma & Bertschinger 1994b).

We find that the best-fit parameter for the $\Omega_\nu = 0.3$ model is $\delta_c \approx 1.8$ for a spherical tophat window function instead of the canonical value 1.68 (Fig. 1). This difference in $\delta_c$ significantly changes the predicted abundance for rare objects at high redshift. (A Gaussian window function requires a smaller $\delta_c$, in this case $\delta_c \approx 1.7$. The tophat gives a somewhat better match in the shape.) The $\Omega_\nu = 0.2$ model requires different $\delta_c$ for early and late epochs. At $z \lesssim 3$, $\delta_c \approx 1.8$ provides a reasonable fit, but this choice underpredicts the halo abundance by a factor of 3 at $z = 5.3$ and 6 at $z = 7.4$. At $z > 3$, $\delta_c = 1.68$ provides a better fit (Fig. 1). This trend of favoring a smaller $\delta_c$ at earlier times was also observed in the pure CDM model by Gelb & Bertschinger (1994). Overall, however, the slope $d\log N/d\log M$ is steeper in the Press-Schechter approximation than in our simulations.

## 3. Observational Constraints

The comoving number density of quasars brighter than $M_B = -26$ from the Palomar grism survey (Schmidt et al. 1991) is plotted in Figure 1. The data point at $z=2.2$ is based on C IV-selected quasars; the points at higher $z$ are from the recent analysis of Ly$\alpha$-selected quasars (Schmidt et al. 1994). The highest-redshift ($z = 4.5$) data point provides the most stringent constraint. For $\Omega_\nu = 0.3$, no massive halos have formed at this redshift in our simulation box, so the theoretical predictions can be estimated only by extrapolating the best-fit Press-Schechter curves to our simulation results. If quasars are active for only a fraction $f$ of the time, then the required number of host halos increases by a factor $f^{-1}$. Even in the most optimistic case ($f = 1$), when 100% of the dark matter halos at high redshift host quasars at any one time, our numerical results show that every halo of mass greater than $\sim 5 \times 10^{11}\ M_\odot$ would have to host a luminous quasar for this model to match the observations. For $\Omega_\nu = 0.2$, this mass rises to $\sim 7 \times 10^{12}\ M_\odot$.

We can assess whether these requirements are plausible by assuming that a quasar corresponds to a massive black hole radiating at the Eddington limit. With a bolometric correction factor of 6.0 (Sanders et al. 1989; Haehnelt & Rees 1993), $M_B < -26$ corresponds to a black hole mass $M > 7 \times 10^8 \, M_\odot$. The black hole forms presumably by collapse of baryons, which in our models make up only 5% of the total mass. Thus, assuming 1% of the baryons go into the hole (a reasonable but uncertain estimate), the minimum galactic mass required to produce such a hole is $1.4 \times 10^{12} \, M_\odot$. In order to account for the observed numbers of quasars, the $\Omega_\nu = 0.3$ model requires host galaxy masses at least 3 times smaller from above. We have assumed that the time lag between galactic halo formation and black hole formation is negligible; relaxing this assumption only increases the difficulty (Turner 1991). A constant quasar space density from $z \approx 2$ to 4 (Irwin et al. 1991) would also enlarge the discrepancy. By contrast, the $\Omega_\nu = 0.2$ model appears compatible with the high-redshift quasar counts. However, we cannot completely eliminate the $\Omega_\nu = 0.3$ model based on quasar abundances because it is possible, in principle, that black holes form quickly and accrete more than 1% of the baryonic mass in galaxies hosting quasars.

An independent test comes from the amount of neutral gas in galaxies associated with damped Ly$\alpha$ systems. Lanzetta (1993) and Lanzetta, Wolfe, & Turnshek (1993) estimated that the fraction of the critical density in the dense neutral gas associated with damped Ly$\alpha$ systems is (for $q_0 = 0.5$) $\Omega_g = (5.8 \pm 1.2) \times 10^{-3} h_{50}^{-1}$ at $3.0 < z < 3.5$, a value that can be compared with predictions of models.

To calculate the amount of dark mass in our simulations that is associated with the observed damped Ly$\alpha$ systems, we first compute for each of our Denmax halos (without any overdensity cut) the mass column density as a function of projected radius from the halo center. We obtain the corresponding hydrogen column density $N(\mathrm{HI})$ by multiplying the dark mass column density by $\Omega_b (\mu m_\mathrm{H})^{-1}$ (where $m_\mathrm{H}$=hydrogen mass; $\mu$=1.33 to take into account the 25% of helium by mass) under the assumption that the gas and dark matter are distributed similarly in the outer parts of the halos. To compare fairly with observations, we then compute for each halo only the baryonic mass (including both hydrogen and helium) in the cylinder whose $N(\mathrm{HI})$ exceeds the corresponding completeness threshold $2 \times 10^{20}$ cm$^{-2}$ in the survey of Lanzetta et al. (1993). Figure 2a shows $\Omega_g$ computed in this way at $z = 3.2$ for halos more massive than $M$. For comparison we show $\Omega_g$ computed assuming that *all* the baryons in halos overdense by factors of 50 and 200 contribute to damped Ly$\alpha$ absorption. Figure 2b compares the column density distribution function $f(N)$ [related to $\Omega_g$ by $\Omega_g = H_0 c^{-1} \mu m_\mathrm{H} \rho_\mathrm{crit}^{-1} \int_{N_\mathrm{min}}^\infty N f(N) dN$] observed by Lanzetta et al. with the predictions of the $\Omega_\nu = 0.2$ model.

The only assumption behind our calculation is that the damped Ly$\alpha$ systems are embedded in dark halos where the gas and dark matter are distributed alike with a constant ratio given by primordial nucleosynthesis. Because the corresponding halo radii exceed 40 kpc for $M > 10^{12} M_\odot$, dissipative effects should be small so that our assumption is reasonable. Alternatively, if all baryons in dark halos dissipate to high column density, the reader may refer to the dotted lines in Figure 2a. Note that the thick lines in Figure 2a in fact give *higher* $\Omega_g$ for $M \gtrsim 10^{11} M_\odot$ than the dotted lines for virialized halos. This is because we conservatively used *all* Denmax halos (and not just those with $\delta\rho/\rho \geq 200$) for the thick lines when we computed the column densities. In either case our approach is more reliable than estimates based on analytic approximations such as the Press-Schechter formula combined with assumptions such as universal velocity-dispersion/halo mass relations. Our procedure likely overestimates $\Omega_g$ because we assume that all hydrogen contributes to the absorption (i.e., there is no ionization and no dark baryons). Star formation and ionization by shocks and ultraviolet background radiation will further reduce our estimate of $\Omega_g$ in the models.

Figure 2a shows the striking discrepancy between the $\Omega_\nu = 0.3$ model and the observation at $z \approx 3.2$: only 0.007% of the total mass density in this model is in the form of high column-density baryonic clouds compared to 0.058% from observations. For $\Omega_\nu = 0.2$, the fraction rises to $\sim 0.02\%$, still a factor $\sim 3$ below the value needed to match the observation, even with our conservative assumptions. The column density distribution predicted by the model is also too steep (Fig. 2b). Ionization and gaseous dissipation will decrease the slope of $f(N)$ but ionization will also decrease $\Omega_g$, making the discrepancy worse. If the power spectrum normalization is increased to $Q_{\rm rms-PS} = 21$ $\mu$K ($2.3\sigma$ above the estimate of Bennett et al. 1994), $\Omega_g$ is increased by a factor $\sim 1.5$, still not enough for $\Omega_\nu = 0.3$ to match the observations. We thus conclude that if damped Ly$\alpha$ systems are indeed embedded in dark halos with a primordial baryon fraction, then their high abundance constrains the neutrino mass density in CDM+HDM models to be $\Omega_\nu \lesssim 0.2$, providing an upper bound of 4.7 eV on the most massive light neutrino.

Klypin et al. (1994) have reached a different conclusion. However, their Figure 2 indicates that even all the galaxies with total mass exceeding $10^{10} M_\odot$ do not provide enough Ly$\alpha$ absorption at $z \approx 3.2$ for $\Omega_\nu \geq 0.25$. Our numerical calculations (Fig. 2) suggest that the damped Ly$\alpha$ absorbers in CDM+HDM models with $\Omega_\nu \geq 0.2$ come from halos more massive than $5 \times 10^{11} M_\odot$.

Our results place very strong constraints on the CDM+HDM models. While the problems of high-redshift galaxy formation can be ameliorated by decreasing the HDM fraction further, this has the effect of increasing the masses and relative velocities of galaxies at $z = 0$. Klypin et al. (1993) and Nolthenius, Klypin, & Primack (1994) concluded that the $\Omega_\nu = 0.3$ model does not produce excessive small-scale power at $z = 0$, but questions remain for models with smaller $\Omega_\nu$. In a later paper we shall present our own results on the properties of $\Omega_\nu = 0.2$ and 0.3 models at $z = 0$.

We are grateful to Maarten Schmidt for providing the current grism survey data prior to publication. We thank Sterl Phinney and Chuck Steidel for helpful comments on the manuscript. C.-P. Ma acknowledges support from a PMA Division Fellowship at Caltech. E. B. acknowledges support from NSF grants AST90-01762 and ASC93-18185. Supercomputer time was provided by the National Center for Supercomputing Applications.

# REFERENCES

Bennett, C. L., et al 1994, preprint COBE 94-01, astro-ph/9401012

Bertschinger, E., & Gelb, J. M. 1991, Computers in Physics, 5, 164

Cen, R., & Ostriker, J. P. 1993, preprint POP-534

Davis, M., Summers, F. J., & Schlegel, D. 1992, Nature, 359, 393

Efstathiou, G., & Rees, M. J. 1988, MNRAS, 230, 5p

Gelb, J. M., & Bertschinger, E. 1994, ApJ, 436, in press

Haehnelt, M. G. 1993, MNRAS, 265, 727

Haehnelt, M.G., & Rees, M. J. 1993, MNRAS, 263, 168

Holtzman, J. A., & Primack, J. R. 1993, ApJ, 405, 428

Hunstead, R. W., Pettini, M., & Fletcher, A. B. 1990, ApJ, 365, 23

Irwin, M., McMahon, R. G., & Hazard, C. 1991, in The Space Distribution of Quasars, ed. D. Crampton (ASP Conf. Ser., 21), 117

Jing, Y. P., Mo, H. J., Börner, G., & Fang, L. Z. 1994, A&A, 284, 703

Kashlinsky, A. 1993, ApJ, 406, L1

Katz, N., Quinn, T., & Gelb, J. M. 1993, MNRAS, 265, 689

Kauffmann, G., & Charlot, S. 1994, preprint astro-ph/9402015

Klypin, A., Borgani, S., Holtzman, J., & Primack, J. R. 1994, preprint astro-ph/9405003

Klypin, A., Holtzman, J. A., Primack, J. R., & Regos, E. 1993, ApJ, 416, 1

Lanzetta, K. M. 1993, in The Environment and Evolution of Galaxies, ed. J. M. Shull & H. A. Thronson, Jr. (Dordrecht: Kluwer), 237

Lanzetta, K. M., Wolfe, A. M., & Turnshek, D. A. 1993, ApJ, submitted

Ma, C.-P., & Bertschinger, E. 1994a, ApJ, 429, 22

Ma, C.-P., & Bertschinger, E. 1994b, ApJ, submitted, preprint astro-ph/9401007

Mo, H. J., & Miralda-Escude, J. 1994, preprint astro-ph/9402014

Nolthenius, R., Klypin, A., & Primack, J. R. 1994, ApJ, 422, L45

Nusser, A., & Silk, J. 1993, ApJ, 411, L1

Press, W. H., & Schechter, P. 1974, ApJ, 187, 425

Sanders, D. B., Phinney, E. S., Neugebauer, G., Soifer, B. T., & Mathews, K. 1989, ApJ, 347, 29

Schaefer, R. K., Shafi, Q., & Stecker, F. W. 1989, ApJ, 347, 575

Schmidt, M., Schneider, D. P., & Gunn, J. E. 1991, in The Space Distribution of Quasars, ed. D. Crampton (ASP Conf. Ser., 21), 109

Schmidt, M., Schneider, D. P., & Gunn, J. E. 1994, in preparation

Steidel, C. C., & Hamilton, D. 1992, AJ, 104, 941

Steidel, C. C., & Hamilton, D. 1993, AJ, 105, 2017

Subramanian, K., & Padmanabhan, T. 1994, preprint astro-ph/9402006

Taylor, A. N., & Rowan-Robinson, M. 1992, Nature 359, 396

Turner, E. L. 1991, AJ, 101, 5

van Dalen, A., & Schaefer, R. K. 1992, ApJ, 398, 33

Walker, T. P., Steigman, G., Schramm, D., Olive, K. A., & Kang, H.-S. 1991, ApJ, 376, 51

Warren, S. J., Hewett, P. C., & Osmer, P. S. 1994, ApJ, 421, 412

Wolfe, A. M., Turnshek, D. A., Lanzetta, K. M., & Oke, J. B. 1992, ApJ, 385, 151

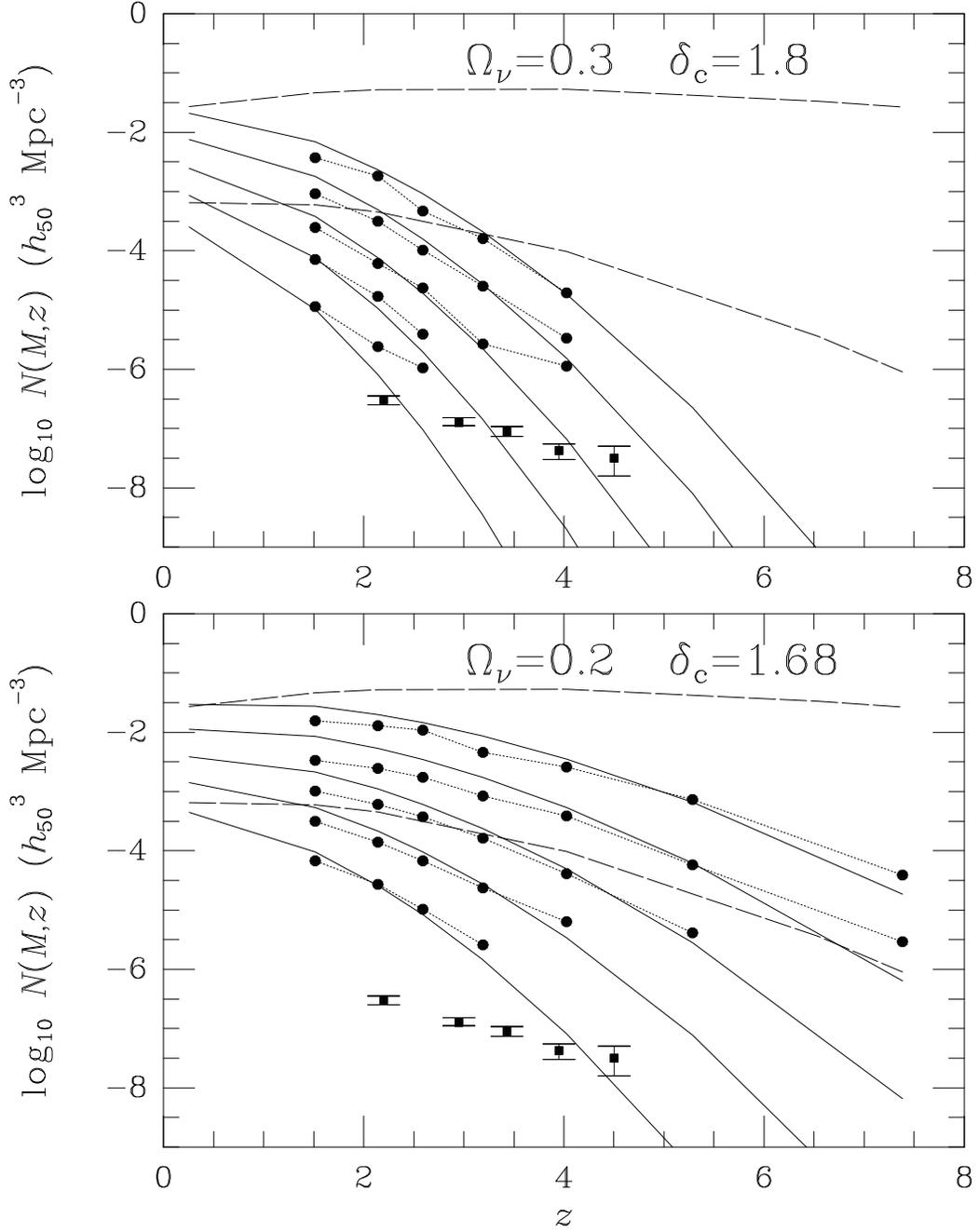

Fig. 1.— The cumulative comoving number density of dark halos as a function of $z$ for the two CDM+HDM models. In each panel, the five solid curves show the Press-Schechter fits with a tophat filter and $\delta_c$ (indicated) above five mass thresholds: $M = 10^{13}$, $10^{12.5}$, $10^{12}$, $10^{11.5}$, and $10^{11} M_\odot$ (from bottom up). The filled circles joined by dotted lines represent the simulated halos above the same mass thresholds. The Press-Schechter function for the CDM model ($\delta_c = 1.8$ for both plots) is also shown for $M > 10^{11}$ (top dashed) and $> 10^{13} M_\odot$ (bottom dashed). The five filled squares with error bars represent the comoving space density of quasars (for $q_0 = 0.5$ and $\Omega = 1$) brighter than $M_B = -26$ from Schmidt et al. (1991, 1994).

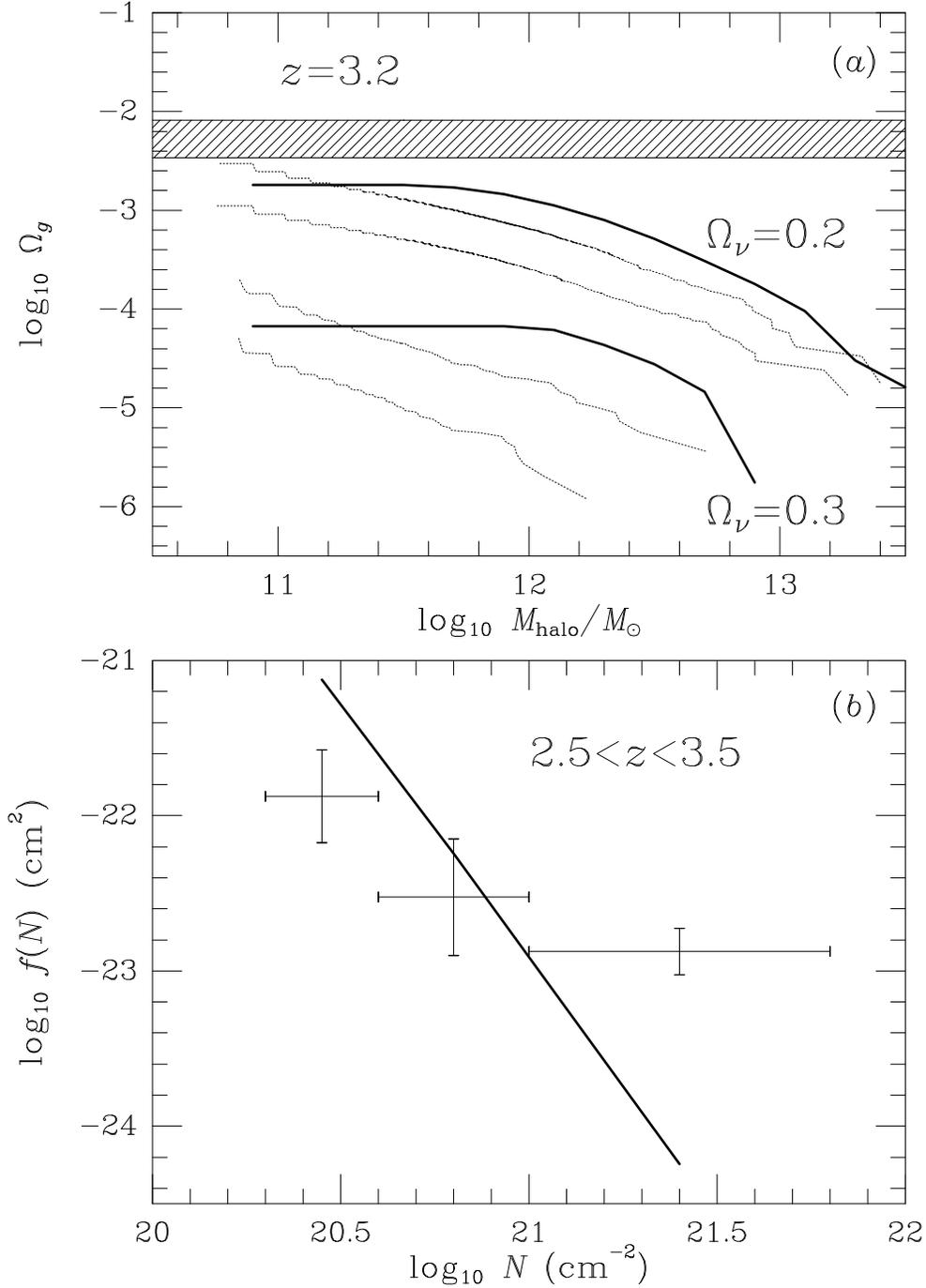

Fig. 2.— (a) The fraction of the critical density in baryons (hydrogen and helium) associated with damped Ly$\alpha$ systems at $z = 3.2$. The thick solid curves (top for $\Omega_\nu$=0.2, bottom for $\Omega_\nu$=0.3) assume the hydrogen column density threshold of Lanzetta et al. (1993). The top two dotted curves show $\Omega_g$ predicted by the $\Omega_\nu$=0.2 model, assuming *all* hydrogen in Denmax halos with $\delta\rho/\rho \geq 200$ (bottom) and $\geq 50$ (top) contributes to Ly$\alpha$ absorption; the bottom two are for $\Omega_\nu$=0.3. The shaded band shows the observed $\Omega_g$ from Lanzetta et al. (b) The column density distribution predicted by the $\Omega_\nu$=0.2 model (thick line) vs. data points from Lanzetta et al. for $2.5 < z < 3.5$.